\documentclass[12pt]{revtex4}
\usepackage{graphicx}
\usepackage{pslatex}

\begin{document}

\title{Melting at the absolute zero of temperature:\\
       Quantum phase transitions in condensed matter}

\author{Matthias Vojta}
\affiliation{Institut f\"ur Theoretische Physik, Universit\"at zu K\"oln,
%Z\"ulpicher Stra{\ss}e 77,
50937 K\"oln, Germany}

\author{Thomas Vojta}
\affiliation{Department of Physics, Missouri University of Science and Technology, Rolla
MO 65409, USA}

\begin{abstract}
{\it
\bigskip
We dedicate this article to our father, Prof. Dr. G\"{u}nter Vojta, on the occasion of
his 80th birthday. While colleagues appreciate him as an excellent scientist with widespread
interests and an amazingly broad knowledge, we see in him the father, teacher,
friend, and much more. Thanks to him, we grew up immersed in the fascinating
world of physics from an early age. His example and encouragement have crucially
influenced our own lives. }

\end{abstract}

\maketitle

%%%%%%%%%%%%%%%%%%%%%%%%%%%%%%%%%%%%%%%%%%%%%%%%%%%%%%%%%%%%%%%%%%%%%%%%%%%
\section{Introduction}
%%%%%%%%%%%%%%%%%%%%%%%%%%%%%%%%%%%%%%%%%%%%%%%%%%%%%%%%%%%%%%%%%%%%%%%%%%%

Phase transitions play an essential role in nature. Everyday transitions include the
boiling of water or the melting of ice, while the transition of a metal into the
superconducting state upon lowering the temperature provides a more complicated example.
The universe itself is thought to have passed through several phase transitions as the
high-temperature plasma formed by the Big Bang cooled to form the world as we know it
today.

Phase transitions occur upon variation of an external control parameter; their common
characteristics is a qualitative change in the system properties. The phase transitions
mentioned so far occur at finite temperature; here macroscopic order (e.g. the crystal
structure in the case of melting) is destroyed by thermal fluctuations. During recent
years, a different class of phase transitions, the so-called quantum phase transitions,
has attracted the attention of physicists. These transitions constitute the subject of this article.

Quantum phase transitions (QPTs) \cite{SGCS97,Sachdev_book99} occur, in a strict sense,
at exactly zero temperature only. A non-thermal control parameter, such as pressure,
magnetic field, or chemical composition, is varied to access the transition point. In
contrast to phase transitions at finite temperature, quantum effects play a decisive
role, as the fluctuations driving the transition follow quantum instead of classical
statistical mechanics. While QPTs were originally considered mere curiosities of
theoretical physics, the impact of this field on modern research is now emerging with
increasing pace.

Violent quantum fluctuations caused by Heisenberg's uncertainty principle are at the
heart of quantum critical points (QCPs) which control continuous QPTs. They affect the
finite-temperature behavior of condensed matter as well, with a multiplicity of new and
unexpected phenomena. For instance, the standard model of metals with electronic
interactions, the Landau Fermi-liquid picture, may break down in the vicinity of a
QPT. In addition, since QPTs occur between nearly degenerate phases whose characteristic
energy scales are driven to zero, small perturbations may become
important which otherwise would be masked by the primary energy scales.
This has the fascinating prospect of inducing novel states of matter around QCPs.

%%%%%%%%%%%%%%%%%%%%%%%%%%%%%%%%%%%%%%%%%%%%%%%%%%%%%%%%%%%%%%%%%%%%%%%%%%%
\section{Example: TlCuCl$_3$ - a coupled dimer magnet}
%%%%%%%%%%%%%%%%%%%%%%%%%%%%%%%%%%%%%%%%%%%%%%%%%%%%%%%%%%%%%%%%%%%%%%%%%%%

To set the stage, we consider as a paradigmatic example the crystalline material
TlCuCl$_3$ which displays several magnetic quantum phase transitions. TlCuCl$_3$ is a
so-called Mott insulator, i.e., it is electrically insulating despite the presence of a
partially filled conduction band. This insulating behavior arises from the strong Coulomb
repulsion which tends to localize electrons in the 3d shells of copper. As a result, each
copper atom in TlCuCl$_3$ carries one unpaired electron with a spin-1/2 magnetic moment
-- these moments are responsible for the variety of magnetic phenomena in TlCuCl$_3$.

\begin{figure}[t]
\centerline{\includegraphics[width=\textwidth]{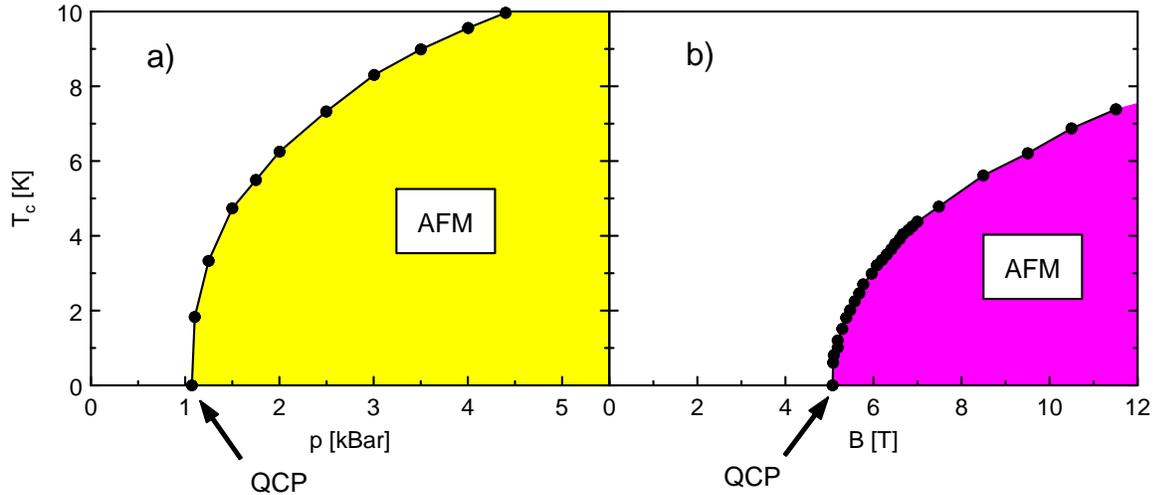}} \caption{Magnetic phase
diagrams of TlCuCl$_3$. a) Pressure--temperature phase diagram at zero magnetic field.
The antiferromagnetic phase breaks the SU(2) spin symmetry of the underlying Heisenberg
model and is bounded by a line of finite-temperature phase transitions. This line
terminates in the quantum critical point (QCP), where antiferromagnetic order can be
tuned by pressure at $T=0$. (Data from Ref.\ \cite{RFSSKGM04}.) b) Field--temperature
phase diagram at ambient pressure. Here, the antiferromagnetic phase breaks a U(1)
symmetry and can be interpreted as Bose-Einstein condensate of magnons. (Data from Ref.\
\cite{OosawaKatoriTanaka01}.) } \label{fig:tlcucl3}
\end{figure}

Fig.~\ref{fig:tlcucl3} shows magnetic phase diagrams of TlCuCl$_3$ as function of
pressure or applied magnetic field and temperature. While the material is a paramagnet at
ambient pressure and low field, it can be driven -- at the lowest temperatures -- into
antiferromagnetic phases either by pressure or field. Conversely, these antiferromagnetic
states at $T=0$ can be destroyed either by increasing temperature -- leading to a
conventional thermal phase transition -- or by lowering pressure or field, thereby
crossing a quantum phase transition point.

\begin{figure}[t]
\includegraphics[width=10cm]{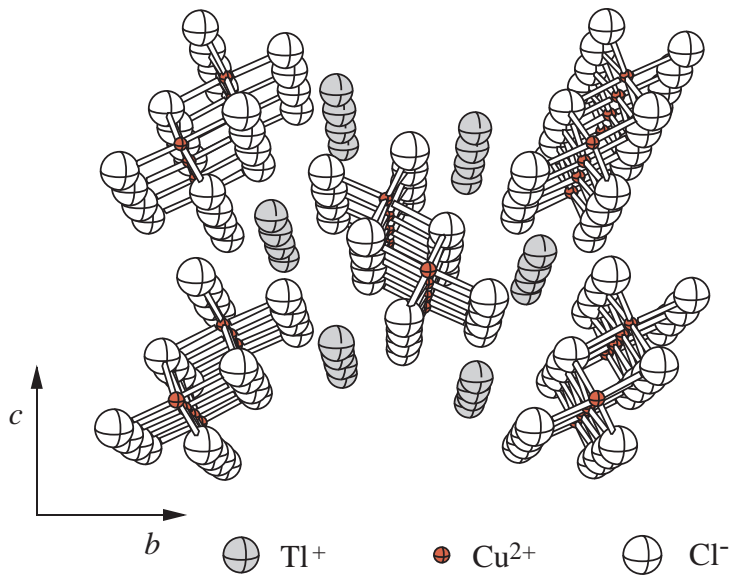}
\caption{ Schematic crystal structure of TlCuCl$_3$. It consists of planar dimers of
Cu$_2$Cl$_6$ which are stacked on top of each other to form infinite double chains along
the crystallographic a axis. Magnetic copper atoms are shown in red.
(Reprinted with permission from Ref. \cite{OTTATSTKK02}.
Copyright 2002 by the American Physical Society.)
}
\label{fig:struct}
\end{figure}

The underlying reason for this interesting behavior is the presence of pairs of spin-1/2
moments, dubbed dimers, which form a three-dimensional network, as shown in
Fig.~\ref{fig:struct}.

Before giving a more detailed description, we collect the basic concepts of phase
transitions and critical behavior \cite{Ma_book76,Goldenfeld_book92} which are necessary
for the later discussions.

%%%%%%%%%%%%%%%%%%%%%%%%%%%%%%%%%%%%%%%%%%%%%%%%%%%%%%%%%%%%%%%%%%%%%%%%%%%
\section{Concepts of classical and quantum phase transitions}
%%%%%%%%%%%%%%%%%%%%%%%%%%%%%%%%%%%%%%%%%%%%%%%%%%%%%%%%%%%%%%%%%%%%%%%%%%%

Phase transitions are traditionally classified into first-order and continuous
transitions. At first-order transitions the two phases co-exist at the transition point,
examples include ice and water at 0$^\circ$ C, or water and steam at 100$^\circ$ C. In
contrast, at continuous transitions, also called critical points, the two phases do not
co-exist, instead they become indistinguishable at the transition point. An important
example is the ferromagnetic transition of iron at 770$^\circ$ C, above which the
magnetic moment vanishes. This phase transition occurs at a point where thermal
fluctuations destroy the regular ordering of magnetic moments -- it happens continuously
in the sense that the magnetization vanishes continuously when approaching the transition
from below.

In the following we concentrate on systems near a continuous phase transition. Such a
transition can usually be characterized by an order parameter; this is a thermodynamic
quantity that is zero in one phase (the disordered) and non-zero and generally non-unique
in the other (the ordered) phase. Very often the choice of an order parameter for a
particular transition is obvious as, e.g., for the ferromagnetic transition where the
total magnetization is an order parameter. However, in some cases finding an appropriate
order parameter is complicated and still a matter of debate, e.g., for the
interaction-driven metal--insulator transition in electronic systems (the Mott transition
\cite{Mott_book90}).

While the thermodynamic average of the order parameter is zero in the disordered phase,
its fluctuations are non-zero. When the critical point is approached, the spatial
correlations of the order parameter fluctuations become long-ranged. Close to the
critical point, their typical length scale, the correlation length $\xi$, diverges as
\begin{equation}
\xi \propto |r|^{-\nu}
\label{divxi}
\end{equation}
where $\nu$ is the correlation length critical exponent and $r$ is some dimensionless
measure of the distance from the critical point. If the transition occurs at a non-zero
temperature $T_c$, it can be defined as $r=|T-T_c|/T_c$.
In addition to the long-range correlations in space there are
analogous long-range correlations of the order parameter
fluctuations in time. The typical time scale for a decay of
the fluctuations is the correlation (or equilibration) time
$\tau_c$. As the critical point is approached the correlation
time diverges as
\begin{equation}
  \tau_c \propto \xi^z \propto |r|^{-\nu z}
\label{divtau}
\end{equation}
where $z$ is the dynamical critical exponent. Close to the critical point, there is no
characteristic length scale other than $\xi$ and no characteristic time scale other than
$\tau_c$. (Note that a microscopic cutoff scale must be present to explain non-trivial
critical behavior, for details see, e.g., Goldenfeld \cite{Goldenfeld_book92}. In a solid
such a scale is, e.g., the lattice spacing.)

The power-law singularities (\ref{divxi}) and (\ref{divtau}) are responsible for the
so-called critical phenomena. At the phase transition point, correlation length and time
are infinite, fluctuations occur on {\em all} length and time scales, and the system is
said to be scale-invariant. As a consequence, all observables depend via power laws on
the external parameters. The set of corresponding exponents -- called critical exponents
-- completely characterizes the critical behavior near a particular phase transition.

One of the most remarkable features of continuous phase transitions is universality,
i.e., the fact that the critical exponents are the same for entire classes of phase
transitions which may occur in very different physical systems. These universality
classes are determined only by the symmetries of the order parameter and by the space
dimensionality of the system. This implies that the critical exponents of a phase
transition occurring in nature can be determined exactly (at least in principle) by
investigating {\em any} simple model system belonging to the same universality class. The
mechanism behind universality is again the divergence of the correlation length. Close to
the critical point the system effectively averages over large volumes rendering the
microscopic details of the Hamiltonian unimportant.

%%%%%%%%%%%%%%%%%%%%%%%%%%%%%%%%%%%%%%%%%%%%%%%%%%%%%%%%%%%%%%%%%%%%%%%%%%%
\section{Quantum phase transitions and the role of quantum mechanics}
%%%%%%%%%%%%%%%%%%%%%%%%%%%%%%%%%%%%%%%%%%%%%%%%%%%%%%%%%%%%%%%%%%%%%%%%%%%

The question of to what extent quantum mechanics is important for understanding a
continuous phase transition has at least two aspects. On the one hand, quantum mechanics
can be essential for understanding the ordered phase, (e.g., superconductivity) -- this
depends on the particular transition considered. On the other hand, one may ask whether
quantum mechanics influences the asymptotic critical behavior. For this discussion we
have to compare two energy scales, namely $\hbar\omega_c$, which is the typical energy of
order parameter fluctuations, and the thermal energy $k_B T$. We have seen in the
preceeding section that the typical time scale $\tau_c$ of the fluctuations diverges as a
continuous transition is approached. Correspondingly, the typical frequency scale
$\omega_c$ goes to zero and with it the typical energy scale
\begin{equation}
  \hbar \omega_c \propto |r|^{\nu z}~.
  \label{eq:energy scale}
\end{equation}
Quantum mechanics will be important as long as this typical energy scale is larger than
the thermal energy $k_B T$; on the other hand, for $\hbar\omega_c \ll k_B T$ a purely
classical description can be applied to the order parameter fluctuations. This implies
that the character of the order parameter fluctuations crosses over from quantum to
classical when $\hbar\omega_c$ falls below $k_B T$.

Now, for any transition occurring at some finite temperature $T_c$, quantum mechanics
will become unimportant for $|r| \lesssim T_c^{1/\nu z}$, in other words, the critical
behavior sufficiently close to the transition is entirely classical. This justifies
calling all finite-temperature phase transitions ``classical''. Quantum mechanics can
still be important on microscopic scales, but classical thermal fluctuations dominate on
the macroscopic scales that control the critical behavior. In contrast, if the transition
occurs at zero temperature as a function of a non-thermal parameter $r$,
the order parameter fluctuations always obey quantum statistical
mechanics. Consequently, transitions at zero temperature are called ``quantum'' phase
transitions. The characteristic scale $\hbar \omega_c \propto r^{\nu z}$ often represents
the energy gap to excitations above the quantum mechanical ground state.

The interplay of classical and quantum fluctuations leads to an interesting phase diagram
in the vicinity of the quantum critical point. Two cases need to be distinguished,
depending on whether or not long-range order can exist at finite temperatures.

\begin{figure}[t]
\centerline{\includegraphics[width=7.5cm,clip]{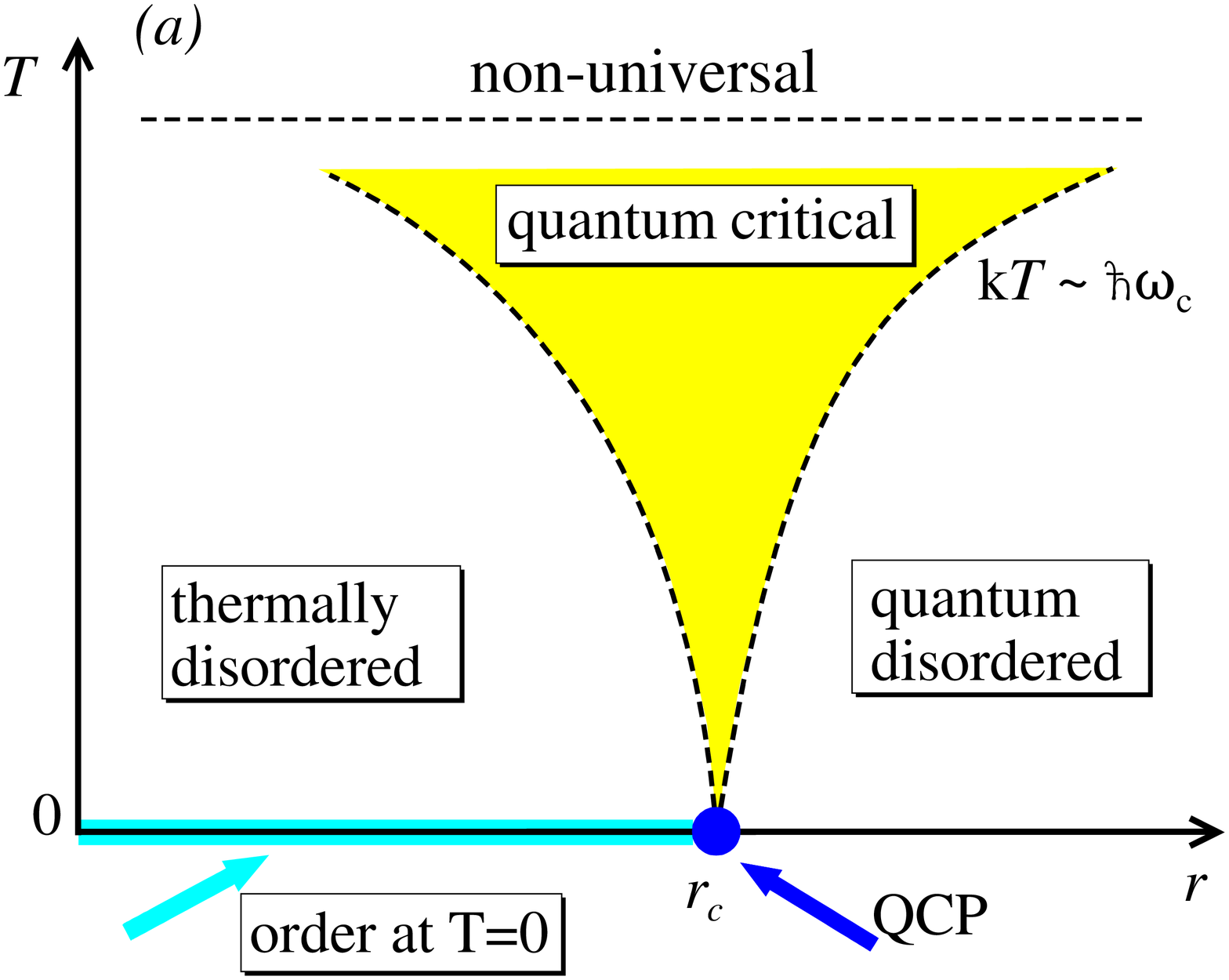}~\includegraphics[width=7.5cm,clip]{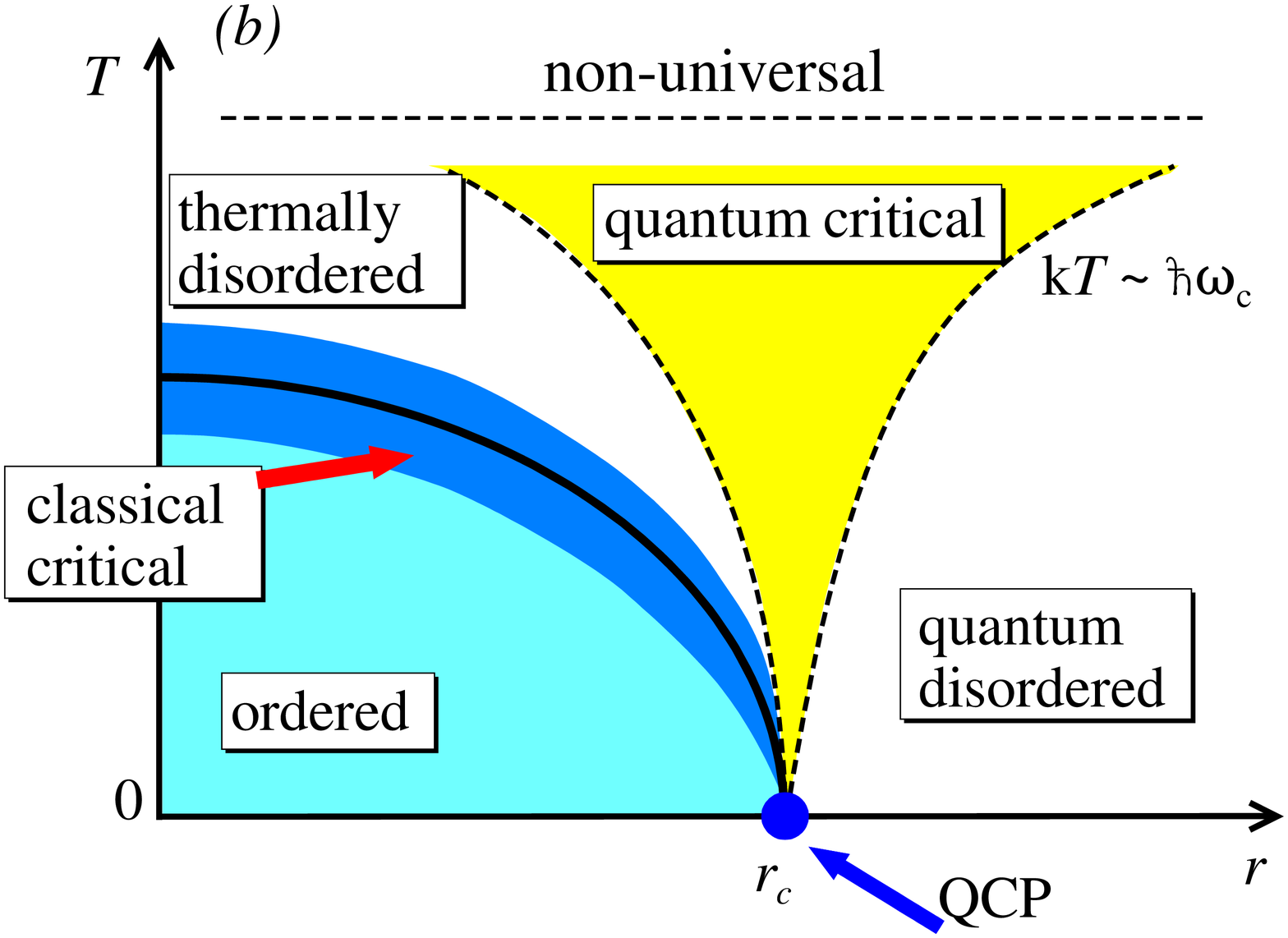}}
\caption[fig1]
{
Schematic phase diagrams in the vicinity of a quantum critical point (QCP).
The horizontal axis represents the control parameter $r$ used to tune
the system through the quantum phase transition, the vertical axis is
the temperature $T$.
a) Order is only present at zero temperature.
The dashed lines indicate the boundaries of the quantum critical region
where the leading critical singularities can be observed;
these crossover lines are given by $k_B T \propto |r-r_c|^{\nu z}$.
b)~Order can also exist at finite temperature.
The solid line marks the finite-temperature boundary between the ordered
and disordered phases. Close to this line, the critical behavior
is classical.
}
\label{fig:qcpvic}
\end{figure}

Fig.~\ref{fig:qcpvic}a describes the situation where order only exists at $T=0$, this is
the case, e.g., in two-dimensional magnets with SU(2) symmetry where order at finite $T$
is forbidden by the Mermin-Wagner theorem. In this case there will be no true phase
transition in any experiment carried out at finite temperature. However, the
finite-$T$ behavior is characterized by three very different regimes, separated by
crossovers. For low $T$ and $r>0$ thermal effects are negligible ($T \ll r^{\nu z}$), and
the critical singularity is cutoff by the deviation of the control parameter $r$ from
criticality. This regime is dubbed ``quantum disordered'' and characterized by
well-defined quasiparticle excitations; for a magnetic transition in a metallic system
this will be the usual Fermi-liquid regime. For $r<0$ and $T>0$, we are in the
``thermally disordered'' regime; here the order is destroyed by thermal fluctuations of
the ordered state (yet quasiparticles are still well defined on intermediate scales). A
completely different regime is the high-temperature regime above the QCP. In this
``quantum critical'' regime \cite{ChakravartyHalperinNelson89}, bounded by crossover
lines $T \sim |r|^{\nu z}$, the critical singularity is cutoff by the finite temperature.
The properties are determined by the unconventional excitation spectrum of the quantum
critical ground state, whose main characteristics is the {\em absence} of conventional
quasiparticle-like excitations, which are replaced by a critical continuum of
excitations. In the quantum critical regime, this continuum is thermally excited,
resulting in unconventional power-law temperature dependencies of physical observables.
Universal behavior is only observable in the vicinity of the quantum critical
point, i.e., when the correlation length is much larger than microscopic length
scales.
Quantum critical behavior is thus cut off
at high temperatures when $k_B T$ exceeds characteristic microscopic energy
scales of the problem -- in magnets this cutoff is, e.g., set by the typical
exchange energy.

If order also exists at finite temperatures, Fig.~\ref{fig:qcpvic}b,
the phase diagram is even richer.
Here, a real phase transition is encountered upon variation of $r$
at low $T$;
the quantum critical point can be viewed as the endpoint of a line of
finite-temperature transitions.
As discussed above, classical fluctuations will dominate in the vicinity of
the finite-$T$ phase boundary, but
this region becomes narrower with decreasing temperature, such that
it might even be unobservable in a low-$T$ experiment.
The fascinating quantum critical region is again at finite temperatures above
the quantum critical point, and the thermally disordered regime is
restricted to temperatures $T>T_c$.

It is instructive to discuss QPTs in terms of
behavior of the low-energy many-particle states of the quantum system upon a change of $r$,
starting with finite system size and then taking the thermodynamic limit.
A first-order quantum phase transition corresponds to a ground-state level crossing at
$r=r_c$, which remains sharp in the thermodynamic limit. In contrast, a continuous quantum phase
transition manifests itself in an avoided level crossing, which becomes sharper upon increasing the system
size. Here, a large number of low-lying excited state approach the ground state,
reflecting the fact that the characteristic scale $\omega_c$ vanishes at $r=r_c$.
In the thermodynamic limit, the avoided crossing becomes sharp, but only the second
derivative of the ground state energy $E(r)$ becomes singular.

%%%%%%%%%%%%%%%%%%%%%%%%%%%%%%%%%%%%%%%%%%%%%%%%%%%%%%%%%%%%%%%%%%%%%%%%%%%
\section{Back to the example: Coupled dimer magnets}
%%%%%%%%%%%%%%%%%%%%%%%%%%%%%%%%%%%%%%%%%%%%%%%%%%%%%%%%%%%%%%%%%%%%%%%%%%%

Let us return to coupled dimer magnets, with TlCuCl$_3$ being a prime example.
The copper atoms carry spin 1/2, with mutual antiferromagnetic exchange interactions
of Heisenberg type.
The monoclinic crystal structure contains two copper atoms per
unit cell, hence the spins are naturally paired into dimers, and the structure can be
understood as three-dimensional arrangement of coupled dimers.

A minimal microscopic model for the relevant magnetic degrees of
freedom is an antiferromagnetic quantum Heisenberg with spatially modulated coupling,
with the Hamiltonian
\begin{equation}
H =
\sum_{\langle ij \rangle} J_{ij} \vec S_i \cdot \vec S_j  -
\vec h \cdot \sum_i \vec S_i~.
\end{equation}
Here $\vec S_i$ is a vector spin operator for the copper spin
at lattice site $i$, $\vec h$ is an external magnetic field.
The couplings $J_{ij}$ obey
\begin{equation}
J_{ij} = \left\{
\begin{array}{ll}
J  & {\rm intra-dimer} \\
J' & {\rm nearest~inter-dimer}
\end{array}
\right. ;
\end{equation}
for simplicity we have reduced the inter-dimer couplings to a single coupling constant.

In the absence of a field, the ground state is determined by the ratio $J'/J$. For $J'
\ll J$, quantum mechanical singlets are formed on each dimer, and the ground state is a
quantum paramagnet. Its elementary magnetic excitations are so-called triplons, i.e.
dispersive spin-1 excitations with a minimum excitation energy $\Delta$. In the limit of
$J'\to 0$ the ground state is simply given by a product state of dimer singlets, and the
gap $\Delta=J$. If, in contrast, $J' \sim J$, the system is a three-dimensional
antiferromagnet displaying magnetic long-range order. (We are assuming that the magnetic
frustration inherent in the lattice structure is small.) Thus, tuning the ratio $J'/J$
will lead to a magnetic order--disorder transition in the ground state. This is exactly
what happens upon applying pressure to TlCuCl$_3$, Fig.~\ref{fig:tlcucl3}a. Increasing
the pressure $p$ changes the lattice constants such that the ratio $J'/J$ increases, and the system is
driven into an antiferromagnetic state, accompanied by a spontaneous breaking of the
SU(2) symmetry of the Heisenberg model. The gap of the paramagnet vanishes upon
approaching the critical pressure $p_c$ as $\Delta\propto(p_c-p)^{\nu z}$ with mean-field critical exponents
$\nu=1/2$ and $z=1$ (with logarithmic corrections).

Applying a field at ambient pressure does not affect the paramagnetic ground state,
but causes a Zeeman splitting of the spin-1 triplon excitations.
At a critical field $H_{c1}$, the gap of the lowest triplon branch closes,
and the system is again driven into an ordered state, Fig.~\ref{fig:tlcucl3}b.
This ``canted'' state has a finite uniform magnetization in the direction of the applied field and a
spontaneous antiferromagnetic order perpendicular to the field direction.
Further increasing the field leads to a second transition at $H_{c2}$ where the
systems enters a fully polarized state.
(In TlCuCl$_3$, $H_{c2}$ is expected to be around 90 T which has not been reached
experimentally.)

We emphasize that the physics of the field-driven case is qualitatively
different from the pressure-driven situation.
In a field, the spin symmetry of the Hamiltonian is reduced from SU(2) to U(1),
and the ordered phase breaks this U(1) symmetry. The lowest triplon
branch of the paramagnet can be interpreted as a bosonic excitation on top of the singlet
ground state. In the field-ordered phase, the ground state can be written as a
superposition of singlet and triplet -- a ``triplon condensate'' -- with a complex ratio
between singlet and triplet component, determining the orientation of the
staggered magnetization perpendicular to the field. Upon raising the temperature, the
antiferromagnetic order is destroyed at $T_N$, but the density of triplons turns out to
change in a non-singular fashion across $T_N$. Remarkably, the finite-temperature
transition at $T_N$ can be understood as Bose-Einstein condensation of triplons (or
magnons): These are pre-existing bosons which acquire phase coherence below $T_N$. The
phase of the Bose condensate is observable as orientation of the staggered magnetization.
In contrast, the triplon density changes singularly upon crossing the quantum
phase transition at $H_{c1}$: The density is simply zero below $H_{c1}$ and finite above.
Technically, this QPT is in the universality class of the dilute Bose gas.

%%%%%%%%%%%%%%%%%%%%%%%%%%%%%%%%%%%%%%%%%%%%%%%%%%%%%%%%%%%%%%%%%%%%%%%%%%%
\section{Conclusions and outlook}
\label{sec:sum}
%%%%%%%%%%%%%%%%%%%%%%%%%%%%%%%%%%%%%%%%%%%%%%%%%%%%%%%%%%%%%%%%%%%%%%%%%%%

This article has illustrated aspects of zero-temperature phase transitions in quantum
systems. Conceptually, quantum phase transitions open a field of fascinating physics, as
they are connected to the peculiar properties of the quantum critical ground state.
Quantum criticality also provides new perspectives in the study of correlated systems,
where intermediate-coupling phenomena are hardly accessible by standard weak- or
strong-coupling perturbative approaches. A promising route starts by identifying quantum
critical points between stable phases, and then uses these as vantage points for
exploring the phase diagram by expanding in the deviation from criticality.

While thermodynamic properties of magnetic quantum phase transitions in insulators, as in
our example, are relatively well understood (this also applies to phase transitions with
other conventional order parameters), there is a variety of challenges in current
research: (i) Quantum phase transitions in metals, often accompanied by non-Fermi liquid
behavior, are less understood, due to the coupling between order parameter fluctuations
and particle-hole excitations of the metal. (ii) Transitions without a local order
parameter are known to exist, but they are a hard task for theory, with examples being
the Mott transition and various types of topological transitions. (iii) The influence on
quantum phase transitions of quenched disorder, which is unavoidable is solids, is very
strong. Even simple models lead to spectacular effects,  but little is known for more
realistic types of disorder. (iv) Transport, and more generally, non-equilibrium
properties, near quantum phase transitions are an exciting and difficult venue of
research. Here, beautiful experiments with ultracold atomic gases have triggered a lot of
interest. It is clear that we have just scratched the surface of this field, and much
exciting progress is expected in the future.

%%%%%%%%%%%%%%%%%%%%%%%%%%%%%%%%%%%%%%%%%%%%%%%%%%%%%%%%%%%%%%%%%%%%%%%%%%

\bigskip

This work was supported by the NSF under Grant No. DMR-0339147, by Research Corporation,
and by the University of Missouri Research Board as well as by the DFG through SFB 608
and FG 960.

%%%%%%%%%%%%%%%%%%%%%%%%%%%%%%%%%%%%%%%%%%%%%%%%%%%%%%%%%%%%%%%%%%%%%%%%%%%

% \newpage
%\section*{References}

\end{document}